\documentclass[amsmath,amssymb,twocolumn]{revtex4}
\usepackage{graphicx}

\def\ri{{\rm i}}

\def\PT{$\mathcal{PT}$}
\def\[{\begin{equation}}
\def\]{\end{equation}}

\begin{document}
\title{Symmetry breaking of solitons in one-dimensional parity-time-symmetric optical potentials}
\author{Jianke Yang}

\address{Department of Mathematics and Statistics, University of Vermont, Burlington, VT 05401, USA}

\begin{abstract}
Symmetry breaking of solitons in a class of one-dimensional
parity-time (\PT) symmetric complex potentials with cubic
nonlinearity is reported. In generic \PT-symmetric potentials, such
symmetry breaking is forbidden. However, in a special class of
\PT-symmetric potentials $V(x)=g^2(x)+\alpha g(x)+ i g'(x)$, where
$g(x)$ is a real and even function and $\alpha$ a real constant,
symmetry breaking of solitons can occur. That is, a branch of
non-\PT-symmetric solitons can bifurcate out from the base branch of
\PT-symmetric solitons when the base branch's power reaches a
certain threshold. At the bifurcation point, the base branch changes
stability, and the bifurcated branch can be stable.
\end{abstract}

\maketitle

Light propagation in optical waveguides is often modeled by
Schr\"odinger-type equations under the paraxial approximation
\cite{Kivshar_book,Yang_book}. If the waveguide contains gain and
loss, the optical potential of the Schr\"odinger equation would be
complex \cite{PT_2005,Christodoulides2007}. A surprising phenomenon
is that if this complex potential satisfies parity-time (\PT)
symmetry, i.e., if the refractive index is even and the gain-loss
profile odd, then the linear spectrum can still be all-real, thus
allowing stationary light transmission
\cite{Bender1998,Ahmed2001,Musslimani2008,Zezyulin2012a}. A number
of other unusual properties have been discovered in \PT-symmetric
systems as well
\cite{Musslimani2008,Guo2009,Segev2010,PT_lattice_exp,Abdullaev2011,coupler1,coupler2,He2012,Longhi_2009,Musslimani_diffraction_2010,Christodoulides_uni_2011,Nixon2012b,Nixon2013,Konotop2012,
Kevrekidis2013, Li2011,
Zezyulin2012b,Wang2011,Lu2011,Hu2011,Zhu2011,Nixon2012,Zezyulin2012a,Dong2012,
Kartashov2013}. One of them is that \PT-symmetric potentials can
support continuous families of \PT-symmetric solitons
\cite{Wang2011,Lu2011,Hu2011,Zhu2011,Nixon2012,Zezyulin2012a,Dong2012},
in contrast with typical dissipative systems where solitons only
exist as isolated solutions \cite{Akhmediev_book}.

However, symmetry breaking of solitons in \PT-symmetric potentials
has never been reported. Based on a perturbative analysis, it was
shown that this symmetry breaking requires an infinite number of
nontrivial conditions to be satisfied simultaneously, which cannot
occur in a generic \PT-symmetric potential \cite{YangStud2014}. From
a similar analysis, it was shown that a generic non-\PT-symmetric
complex potential could not support continuous families of solitons
either \cite{YangPhysLett2014}. Recently, a continuous family of
solitons was reported in a special non-\PT-symmetric potential
$V(x)=g^2(x)+ i g'(x)$, where $g(x)$ was a real asymmetric localized
function \cite{Tsoy2014}. This is a surprising result, and it
suggests that symmetry breaking of solitons might also be possible
in special types of \PT-symmetric potentials.

In this paper, we show that for the special type of \PT-symmetric
potentials $V(x)=g^2(x)+\alpha g(x)+ i g'(x)$, where $g(x)$ is an
arbitrary real and even function and $\alpha$ an arbitrary real
constant, symmetry breaking of solitons can indeed occur.
Specifically, in this class of complex potentials, a branch of
non-\PT-symmetric solitons can bifurcate out from the base branch of
\PT-symmetric solitons when the base branch's power reaches a
certain threshold. But if the \PT potential deviates from this
special form, then symmetry breaking generically disappears.
Symmetry breaking of solitons in this class of \PT-symmetric
potentials is a surprising phenomenon, since infinitely many
nontrivial conditions in \cite{YangStud2014} are miraculously
satisfied simultaneously, which makes this symmetry breaking
possible.

The model for nonlinear propagation of light beams in
one-dimensional complex optical potentials is taken as
\begin{equation} \label{Eq:NLS}
\ri \Psi_z + \Psi_{xx} + V(x)\Psi + \sigma |\Psi|^2 \Psi = 0,
\end{equation}
where $z$ is the propagation direction, $x$ is the transverse
direction, and $\sigma=\pm 1$ is the sign of cubic nonlinearity. The
complex potential $V(x)$ is assumed to be \PT-symmetric, i.e.,
\begin{equation} \label{e:PPTcondition}
V^*(x)=V(-x),
\end{equation}
where the asterisk `*' represents complex conjugation.

Solitons in Eq. (\ref{Eq:NLS}) are sought of the form
\[
\Psi(x, z)=\psi(x)e^{i\mu z},
\]
where $\mu$ is a real propagation constant, and $\psi(x)$ is a
localized function which satisfies the equation
\[  \label{e:psi}
\psi_{xx}+V(x)\psi+\sigma |\psi|^2\psi=\mu \psi.
\]

For a generic \PT-symmetric potential $V(x)$, its linear spectrum,
i.e., the set of eigenvalues $\mu$ for the linear Schr\"odinger
equation
\[ \label{e:Schrodinger}
\psi_{xx}+V(x)\psi=\mu \psi
\]
can be all-real
\cite{Bender1998,Ahmed2001,Musslimani2008,Zezyulin2012a}. In
addition, such a potential often admits continuous families of
\PT-symmetric solitons
\cite{Wang2011,Lu2011,Hu2011,Zhu2011,Nixon2012,Zezyulin2012a,Dong2012}.
However, it is very hard for a \PT potential to admit families of
non-\PT-symmetric solitons. Indeed, such solution families have
never been reported to the author's best knowledge. The analytical
reason is that in order for such non-\PT solution families to exist
in a \PT potential, infinitely many nontrivial conditions have to be
met, which is impossible in generic \PT-symmetric potentials
\cite{YangStud2014}.

In this paper, we consider a special class of \PT-symmetric
potentials
\[  \label{e:VPT}
V(x)=g^2(x)+\alpha g(x)+ i g'(x),
\]
where $g(x)$ is an arbitrary real and even function, and $\alpha$ is
an arbitrary real constant. This form of the potential contains that
used in \cite{Tsoy2014} as a special case (when $\alpha=0$). For
certain functions of $g(x)$, the linear spectrum of this potential
can be analytically shown to be all-real \cite{Wadati2008}. But this
all-real spectrum actually holds for many more functions of $g(x)$
since the potential (\ref{e:VPT}) is \PT-symmetric. If $\alpha=0$,
the Schr\"odinger eigenvalue problem (\ref{e:Schrodinger}) for the
potential (\ref{e:VPT}) can be transformed to the Zakharov-Shabat
eigenvalue problem \cite{Tsoy2014,Wadati2008}. If $\alpha\ne 0$, the
Schr\"odinger eigenvalue problem for this potential can be
transformed to a generalized Zakharov-Shabat eigenvalue problem
\cite{Wadati2008}.

For this class of potentials (\ref{e:VPT}), we will show that
continuous families of non-\PT-symmetric solitons can appear through
symmetry-breaking bifurcations. In real symmetric potentials,
symmetry breaking of solitons is a well-known phenomenon
\cite{Malomed_book}. But in complex \PT-symmetric potentials, such
symmetry breaking is very novel. We will numerically demonstrate
this symmetry breaking by a number of examples. Comparison with the
analytical conditions for symmetry breaking in \cite{YangStud2014}
will also be made.

\begin{figure}
\begin{center}
\includegraphics[height = 2.5in]{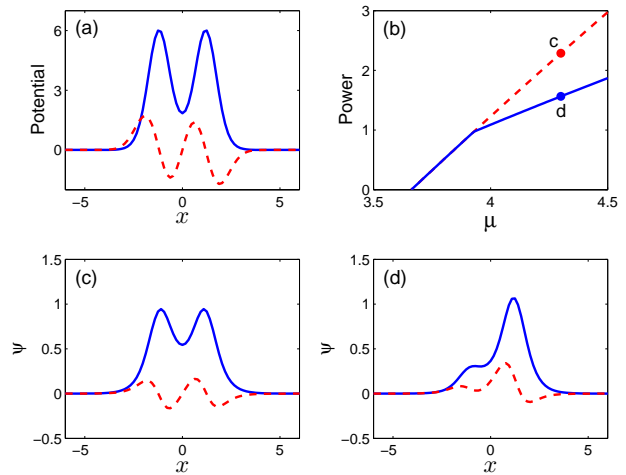}
\caption{Symmetry breaking of solitons in the \PT-symmetric potential (\ref{e:VPT}) for a double-hump function (\ref{e:doublewell})
with $A=2$, $x_0=1.2$, $\alpha=1$ and $\sigma=1$. (a) Profile of the potential $V(x)$ [solid blue: Re($V$); dashed red: Im($V$)].
(b) Power diagram (solid blue: stable branches;
dashed red: unstable branch). (c) \PT-symmetric soliton at point `c' of the power diagram. (d) Non-\PT-symmetric soliton at point `d' of the power diagram.
In (c,d), solid blue is Re($\psi$), dashed is red Im($\psi$), and $\mu=4.3$.                         \label{f:fig1}}
\end{center}
\end{figure}

We first consider a localized double-hump function of $g(x)$,
\[ \label{e:doublewell}
g(x)=A \left(e^{-(x+x_0)^2}+e^{-(x-x_0)^2}\right),
\]
with $A=2$, $x_0=1.2$ and $\alpha=1$. The corresponding \PT
potential $V(x)$ from (\ref{e:VPT}) is plotted in
Fig.~\ref{f:fig1}(a). The linear spectrum of this potential is
all-real, and it contains three positive discrete eigenvalues, the
largest being $3.6614$. From this largest discrete eigenmode, a
family of \PT-symmetric solitons bifurcates out. Under focusing
nonlinearity ($\sigma=1$), the power curve of this solution family
is shown in Fig.~\ref{f:fig1}(b), and the soliton profile at the
marked point `c' (with $\mu=4.3$) is displayed in
Fig.~\ref{f:fig1}(c). Here the soliton's power is defined as
$P=\int_{-\infty}^\infty |\psi(x; \mu)|^2 dx$. What is interesting
is that, at the propagation constant $\mu_c\approx 3.9287$ of this
base power branch, a family of non-\PT-symmetric solitons bifurcates
out. The power curve of this non-\PT-symmetric family is also shown
in Fig.~\ref{f:fig1}(b). At the marked point `d' of the bifurcated
power branch, the non-\PT-symmetric solution is displayed in
Fig.~\ref{f:fig1}(d). It is seen that most of the energy in this
soliton resides on the right side of the potential. In order to
ascertain these non-\PT-symmetric solitons are true solutions to Eq.
(\ref{e:psi}), we have computed them using the
Newton-conjugate-gradient method \cite{Yang_book} and 32 significant
digits (in Matlab with a multiprecision toolbox). These solutions
are found to satisfy Eq. (\ref{e:psi}) to an accuracy of $10^{-30}$,
confirming that they are indeed true solutions.

Since Eq. (\ref{e:psi}) is \PT-symmetric, if $\psi(x)$ is a
solution, so is $\psi^*(-x)$. Thus for each of the non-\PT-symmetric
solitons $\psi(x; \mu)$ in Fig.~\ref{f:fig1}(b), there is a
companion soliton $\psi^*(-x; \mu)$ whose energy resides primarily
on the left side of the potential. In other words, the bifurcation
in Fig.~\ref{f:fig1}(b) is a pitchfork-type symmetry-breaking
bifurcation. This bifurcation resembles that in conservative systems
with real symmetric potentials, which is remarkable since the
present potential is dissipative (with gain and loss).

Linear stability of these solitons can be determined by perturbing
them with normal modes, $\Psi(x,t)=e^{i\mu
z}\left[\psi(x)+q(x)e^{\lambda z}+r^*(x)e^{\lambda^*z}\right]$,
where $q, r \ll \psi$. Inserting this perturbation into Eq.
(\ref{Eq:NLS}), a linearized eigenvalue problem for $(q, r)$ can be
derived, with $\lambda$ being the eigenvalue. The set of all
eigenvalues $\lambda$ is the linear-stability spectrum. If this
spectrum contains eigenvalues with positive real parts, then the
soliton is linearly unstable. Otherwise it is linearly stable.

\begin{figure}
\begin{center}
\includegraphics[height = 1.25in]{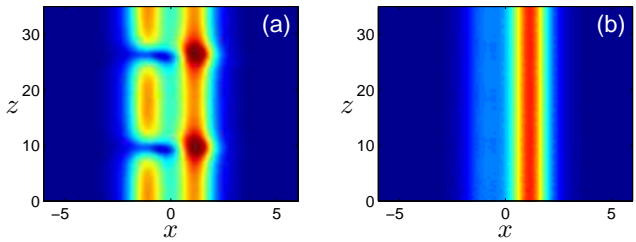}
\caption{(a,b) Nonlinear evolutions of the two solitons in Fig.~\ref{f:fig1}(c,d) under $1\%$ random-noise
perturbations.  \label{f:fig2}}
\end{center}
\end{figure}

We have determined the linear-stability spectra of these solitons by
the Fourier collocation method \cite{Yang_book}. We found that the
base branch of \PT-symmetric solitons is stable before the
bifurcation point ($\mu<\mu_c$). After the bifurcation point, this
base branch becomes unstable due to the presence of a real positive
eigenvalue. However, the bifurcated branch of non-\PT-symmetric
solitons is stable. These stability results are marked on the power
diagram of Fig.~\ref{f:fig1}(b). To corroborate these
linear-stability results, we perturb the two solitons in
Fig.~\ref{f:fig1}(c,d) by $1\%$ random-noise perturbations, and
their nonlinear evolutions are displayed in Fig.~\ref{f:fig2}(a,b).
It is seen from panel (a) that the \PT-symmetric soliton in
Fig.~\ref{f:fig1}(c) breaks up and becomes non-\PT-symmetric. Upon
further propagation, the solution bounces back to almost
\PT-symmetric again, followed by another breakup. In contrast, panel
(b) shows that the non-\PT-symmetric soliton in Fig.~\ref{f:fig1}(d)
is stable against perturbations.

\begin{figure}
\begin{center}
\includegraphics[height = 1.25in]{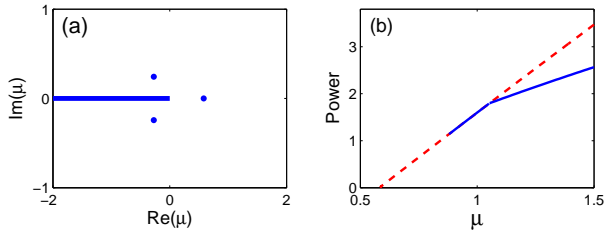}
\caption{Symmetry breaking of solitons in the \PT-symmetric potential (\ref{e:VPT})-(\ref{e:doublewell}) above the phase transition point
(with $A=2$, $x_0=1.2$, $\alpha=-0.9$ and $\sigma=1$). (a) Linear spectrum of the potential (\ref{e:VPT}). (b) Power diagram of symmetry-breaking bifurcations
(solid blue: stable solitons; dashed red: unstable solitons).  \label{f:fig3}}
\end{center}
\end{figure}

For the \PT-symmetric potential (\ref{e:VPT}) with $g(x)$ given by
(\ref{e:doublewell}), we have also tested many other values of $A,
x_0$ and $\alpha$, and observed symmetry-breaking bifurcations as
well. Interestingly, we found that symmetry breaking can still occur
even if the potential is above the phase-transition point, i.e.,
when the linear spectrum of the potential is not all-real
\cite{Bender1998,Ahmed2001,Musslimani2008}. For instance, with the
same $A$ and $x_0$ values as above, when $\alpha=-0.9$, the linear
spectrum of the resulting potential (\ref{e:VPT}) is displayed in
Fig.~\ref{f:fig3}(a). This spectrum contains a pair of complex
eigenvalues, indicating that this \PT-symmetric potential is above
the phase-transition point. This spectrum also contains a discrete
real eigenvalue $\mu\approx 0.5817$, from which a continuous family
of \PT-symmetric solitons bifurcates out. The power curve of these
\PT-symmetric solitons (for $\sigma=1$) is shown in
Fig.~\ref{f:fig3}(b). The low-power segment of this solution branch
is unstable since the potential is above the phase-transition point.
However, when the power is above 1.14, these \PT-symmetric solitons
become stable. At the power value of approximately 1.8, symmetry
breaking occurs, where a branch of non-\PT-symmetric solitons
bifurcates out. Meanwhile the base branch of \PT-symmetric solitons
loses stability. The presence of symmetry breaking and existence of
stable non-\PT-symmetric solitons in a \PT-symmetric potential above
phase transition is remarkable.

\begin{figure}
\begin{center}
\includegraphics[height = 2.5in]{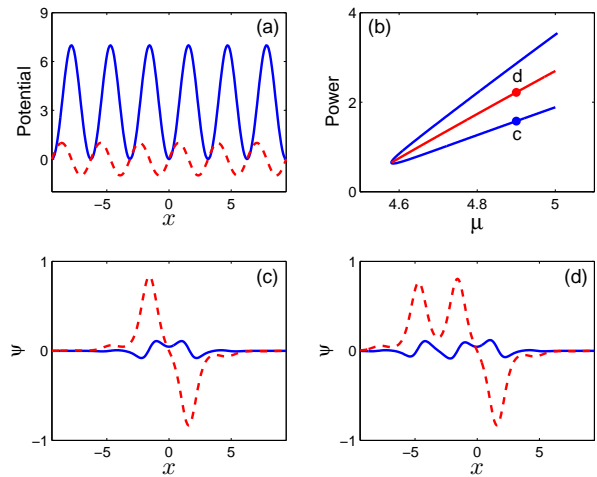}
\caption{Symmetry breaking of solitons in a periodic \PT-symmetric potential (\ref{e:VPT}) with $g(x)$ given by Eq. (\ref{e:gperiod}), $\alpha=6$ and $\sigma=1$.
(a) Profile of the potential $V(x)$ [solid blue: Re($V$); dashed red: Im($V$)].
(b) Power diagram. (c) \PT-symmetric soliton at point `c' of the power diagram. (d) Non-\PT-symmetric soliton at point `d' of the power diagram.
In (c,d), solid blue is Re($\psi$) and dashed red is Im($\psi$).  \label{f:fig4}}
\end{center}
\end{figure}

In addition to the double-hump function (\ref{e:doublewell}),
symmetry breaking can occur in the \PT-symmetric potential
(\ref{e:VPT}) for many other types of real and even functions of
$g(x)$. To demonstrate, we now choose a periodic function for
$g(x)$,
\[ \label{e:gperiod}
g(x)=\sin^2x,
\]
and take $\alpha=6$. The corresponding periodic \PT-symmetric
potential (\ref{e:VPT}) is displayed in Fig.~\ref{f:fig4}(a). In
this potential under focusing nonlinearity ($\sigma=1$), a family of
\PT-symmetric ``dipole" solitons exists in the semi-infinite gap.
The power curve of this ``dipole" family is plotted in
Fig.~\ref{f:fig4}(b), and the profile of the \PT-symmetric soliton
at point `c' of this power curve is shown in Fig.~\ref{f:fig4}(c).
At $\mu_c\approx 4.5801$ of this \PT-symmetric power curve, a family
of non-\PT-symmetric solitons bifurcates out. Its power curve is
also shown in Fig.~\ref{f:fig4}(b) (the middle curve), and the
non-\PT-symmetric soliton at point `d' of this power branch is
displayed in Fig.~\ref{f:fig4}(d).

In a \PT-symmetric potential, symmetry breaking requires infinitely
many nontrivial conditions to be satisfied simultaneously
\cite{YangStud2014}. Due to such stringent conditions, symmetry
breaking cannot occur in a generic \PT-symmetric potential. But for
the special class of potentials (\ref{e:VPT}), we find that those
sequences of conditions are miraculously satisfied. For instance,
for the potential in Fig.~\ref{f:fig1}, we have numerically checked
the first two of those conditions, the first being Eq. (69) in Ref.
\cite{YangStud2014}, and the second being mentioned but not
presented in \cite{YangStud2014}. We verified that those two
(nontrivial) conditions are indeed met. Why potentials (\ref{e:VPT})
satisfy all those infinite conditions is an interesting question
which merits further investigation.

The special nature of potentials (\ref{e:VPT}) is also reflected in
the fact that, when the \PT potential $V(x)$ deviates from those
special forms, symmetry breaking generically disappears. To
demonstrate, we introduce a real parameter $\beta$ into those
potentials,
\[  \label{e:VPT2}
V(x)=g^2(x)+\alpha g(x)+ i \beta g'(x).
\]
Then when $\beta$ moves away from 1, we cannot find symmetry
breaking anymore. For instance, from the potentials in Figs. 1-3,
when we change $\beta$ from 1 to 0.9, we find that families of
non-\PT-symmetric solitons in those figures no longer exist.

It should be pointed out that even though the potentials
(\ref{e:VPT}) are very special among all \PT-symmetric potentials,
they still represent a large class of potentials since $g(x)$ is an
arbitrary real even function, and $\alpha$ is an arbitrary real
constant. In addition, even though we chose a cubic nonlinearity in
our model (\ref{Eq:NLS}), we have found that other types of
nonlinearities (such as cubic-quintic nonlinearity) admit symmetry
breaking as well inside the class of potentials (\ref{e:VPT}). Thus,
by changing different forms of functions $g(x)$ and nonlinearities,
we may still get a wide variety of symmetry-breaking bifurcations.
Whether there are additional types of \PT-symmetric potentials that
admit symmetry-breaking bifurcations is an interesting open
question. Extension of this symmetry breaking to higher spatial
dimensions is another important direction.

In summary, we have reported symmetry breaking bifurcations in a
class of \PT-symmetric potentials (\ref{e:VPT}). From these
bifurcations, families of stable non-\PT-symmetric solitons can be
generated. These results open the door for the study of symmetry
breaking in \PT-symmetric potentials under more general
circumstances.

The author thanks Prof. V. Konotop  and Dr. D. Zezyulin for bringing
Ref. \cite{Tsoy2014} to his attention. This work is supported in
part by AFOSR and NSF.

\end{document}